\documentclass[structabstract]{aa} 

\usepackage[dvips]{graphicx}
\usepackage{amssymb}
\usepackage{xcolor} 
\usepackage{soul} 
\usepackage[normalem]{ulem}
\usepackage{float}
\usepackage{ragged2e}
\usepackage{caption}
\usepackage{subcaption}
\usepackage{rotating}
\usepackage{psfrag}
\usepackage{graphicx}
\usepackage{txfonts}
\usepackage{atbegshi}
\AtBeginDocument{\AtBeginShipoutNext{\AtBeginShipoutDiscard}}
\usepackage{natbib}
\bibpunct{(}{)}{;}{a}{}{,}

\def\mpc{{h^{-1} \rm Mpc}}

\def\kms{{\rm km s^{-1}}}

\begin{document}
   \titlerunning{Unveiling a new structure behind the Milky Way}
   \authorrunning{Galdeano et al.}
   
   \title{Unveiling a new extragalactic structure hidden by the Milky Way}.
   \author{Daniela Galdeano\inst{1}, 
          Gabriel A. Ferrero\inst{2,3},
          Georgina Coldwell\inst{1},
          Fernanda Duplancic\inst{1},\\
          Sol Alonso\inst{1},
          Rogerio Riffel\inst{4}
          \and Dante Minniti\inst{5,6}
          }
   \institute{Departamento de Geof\'{i}sica y Astronom\'{i}a, CONICET, Facultad de Ciencias Exactas, F\'{i}sicas y Naturales, Universidad Nacional de San Juan, Av. Ignacio de la Roza 590 (O), J5402DCS, Rivadavia, San Juan, Argentina             
   \and  
    Facultad de Ciencias Astron\'omicas y Geof\'isicas, Universidad Nacional de La Plata,  Paseo del Bosque s/n, B1900FWA, La Plata, Argentina
    \and
    Instituto de Astrofísica de La Plata, CONICET--UNLP, Paseo del Bosque s/n, B1900FWA, La Plata, Argentina
    \and
    Departamento de Astronomia, Universidade Federal do Rio Grande do Sul, Av. Bento Gon\c{c}alves 9500, 91501-970, Porto Alegre, RS, Brasil
    \and
    Instituto de Astrof\'isica, Facultad de Ciencias Exactas, Universidad Andres Bello, Av. Fernandez Concha 700, Las Condes, Santiago, Chile
    \and
    Vatican Observatory, V00120 Vatican City State, Italy
    }
   \date{Received xxx; accepted xxx}

\abstract
   {The ZOA does not allow clear optical observations of extragalactic sources behind the Milky Way due to  the meaningful extinction of the optical emission of these objects. The observations in NIR wavelengths represent a potential source of astronomical discoveries supporting the detection of new galaxies, completing the picture of the large scale structure in this still little explored area of the sky.}
   {Our aim is to decipher the nature of the overdensity  located behind the Milky Way, in the tile b204 of the VVV survey. }
   {We studied an area of six arcmin around a galaxy concentration located at $l=354.82\degr$ and $b=-9.81\degr$. We selected five galaxies taking into account the source distribution on the sky, in order to optimise the requested time for the observations, and we obtained the spectra with Flamingos 2 long-slit spectrograph at Gemini South 8.1-meter telescope. To identify and characterise the absorption features we have fitted the galaxies underlying spectrum using the {\textsc{starlight}} code together with the IRTF stellar library. 
In addition, the spectroscopic findings are reinforced using complementary photometric techniques such as red-sequence and photometric redshift estimation.}
   {The mean spectroscopic redshift estimated from the NIR spectra is  $z=0.225 \pm 0.014$. This value presents a good agreement with that obtained from photometric analysis, $photoz=0.21 \pm 0.08$, and the probability distribution function of the galaxies in the studied region. Also, the red-sequence slope is consistent with the one expected for NIR observations of galaxy clusters. }
  { The redshifts obtained from both, photometric and spectroscopic techniques are in good agreement allowing the confirmation of the nature of this structure at $z=0.225 \pm 0.014$, unveiling a new galaxy cluster, VVVGCl-B J181435-381432, behind the Milky Way bulge.}
 

\maketitle

%
\section{Introduction}
\label{intro}

Observing extragalactic sources beyond the Milky Way implies 
a continuing challenge since the observations are hampered by the Galactic dust absorption. In this area of the sky, called the zone of avoidance (ZOA), the dust and stars therein obstruct optical observations and a lack of information is produced by dust absorption presenting an incomplete picture of the existing galaxies, and therefore of the extragalactic structures behind the ZOA.

To obtain better information of this region several galaxy catalogues have been developed in different wavelengths. The optical catalogues of \cite{kra00} and \cite{wou04} have allowed the detection of new galaxies at low Galactic latitudes, although the gathering of information is restricted by Galactic dust and stars. Furthermore, near-infrared (NIR), X-ray and \ion{H}{I} radio surveys \citep{Roman1998, ebe02, Vauglin2002, kor04, pat05, skr06, Huchra2012} have detected galaxies and galaxy clusters at low Galactic latitudes. Also, \cite{Jarrett2000} have identified and extracted  extended sources from the Two Micron All-Sky Survey (2MASS) catalogue and \cite{mac19} presented redshifts for 1041 2MASS Redshift Survey galaxies, that previously lacked this information, mostly located within the ZOA.

Moreover, several works have attempted to reveal the presence of 
extragalactic structures, such as groups or clusters, behind the Milky Way. In this line,  \cite{nag04} performed a deep NIR survey with 19 confirmed galaxies and 38 galaxy candidates in a region of $36 \times 36$ arcmin$^{2}$ centred on the giant elliptical radio galaxy PKS 1343–601, on the core of an unknown rich cluster located in the Great Attractor region. Also, \cite{ske09} published a deep $K\!s$ band photometric catalogue containing 390 sources (235 galaxies and 155 galaxy candidates), in a region of $45 \times 45$ arcmin$^{2}$, around the core of the rich nearby Norma cluster (ACO3627).

Furthermore, using the \ion{H}{I} Parkes All-Sky Survey, \cite{sta16} have observed 883 galaxies, delineating possible clusters and superclusters in the Great Attractor region, at low Galactic latitude. On the other hand, \cite{schr19} published a catalogue with 170 galaxies, from the blind \ion{H}{I} survey with the Effelsberg 100m radio telescope, located in the northern region of the ZOA. These new large-scale extragalactic structures could be part of possible filaments at the edge of the local volume. Also, \cite{kra17} discovered a supercluster of galaxies in the ZOA by performing  optical spectroscopic observations of galaxies on the Vela region. These results imply a great advance but still leaving a large unexplored area.

The NIR public survey VISTA Variables in V\'ia L\'actea (VVV; \citep{min10,sai12}) has proven that although the main scientific goals of VVV are related to stellar sources \citep{min11,bea13,iva14}, its exquisite depth (3 magnitudes deeper than 2MASS) and angular resolution make it an excellent tool to find and study extragalactic objects in the ZOA. For example, \cite{amo12} have identified 204 new galaxy candidates from the VVV photometry of a 1.636 square degrees region near the Galactic plane, increasing by more than an order of magnitude the surface density of known galaxies behind the Milky Way. Further, \cite{bar18} found 530 new galaxy candidates in two tiles in the region of the Galactic disk using a combination of SExtractor and PSFEx techniques to detect and characterise these candidates, and \cite{bar21} presented the VVV NIR galaxy catalogue containing 5563 galaxies beyond the MW disk.

Related to extragalactic structures, \cite{col14} have found the VVV NIR galaxy counterparts of a new cluster of galaxies at redshift $z = 0.13$ observed in X-ray with SUZAKU \citep{mor13}. They detected 15 new candidate galaxy members, within the central region of the cluster up to 350 kpc from the X-ray peak emission, with typical magnitudes and colours of galaxies. In addition, \cite{bar19} confirmed the existence of the first galaxy cluster, discovered by the VVV survey beyond the galactic disk, by using spectroscopic data from the spectrograph Flamingos 2 (hereafter F2) at  Gemini South Observatory. More recently, \cite{gal22} present an NIR view of Ophiuchus, the second brightest galaxy cluster in the X-ray sky, finding seven times more cluster galaxy candidates than the number of reported Ophiuchus galaxies in previous works.  

In \citet[][here after G21]{gal21} we found an unusual concentration of galaxies by exploring the $b204$ VVV tile, located at low latitude in the bulge region. In this area, of 1.636 square degrees, we detect 624 extended sources where 607 correspond to new galaxy candidates that have been catalogued for the first time. By exploring the spatial galaxy distribution we found a  smaller region with  a  radius of 15 arcmin with a noticeable higher density, representing approximately 12\% of the whole tile $b204$. This region contains $ 118$ visually confirmed galaxies, three times higher density than the remaining tile area. Also, the comparison of the number of galaxies in this area  with the values obtained from mock catalogues allowed to reinforce the existence of a galaxy overdensity. Even though the results from G21 provide strong evidence of extragalactic structures, the confirmation throughout  spectroscopic redshifts measuring ensures a better knowledge of the nature of this structure. Based on these studies, in this paper we present the results of Gemini South Observatory spectroscopic NIR observations of galaxies from the detected overdensity in G21. Moreover we complement the analysis with systematic photometric studies of this region with the aim of unveil new extragalactic structures behind the ZOA.

This paper is structured as follows: in \S \ref{re2} we describe the data and the galaxy sample used for the analysis. Also, the spectroscopic observations and data reduction are presented in this section. Afterwards, in \S \ref{resN} we present our general results, describing in \S \ref{SR} the techniques to obtain spectroscopic redshift. Afterwards, the red-sequence analysis is presented in \S \ref{RS}, and in \S \ref{PR} we describe the method to obtain photometric redshift. Finally, we discuss and summarise our main conclusions in \S \ref{sum}. The adopted cosmology throughout this paper is $\Omega = 0.3$, $\Omega_{\Lambda} = 0.7$, and $H_0 = 100~ \kms \rm Mpc$.

\section{The data}
\label{re2}

The results obtained from G21 motivated us to perform a deeper research of the reported overdensity area of the sky. With this aim we selected galaxies to carry out NIR spectroscopy and decipher the nature of this structure. In order to optimise the requested time for the observations we take into account the source distribution on the sky. Then, we choose five galaxies to perform the spectroscopic confirmation, by using two F2 slit positions to observe two and three galaxies simultaneously in each position. In addition, the selected galaxies are bright enough to guaranteed a suitable S/N, reducing the observation time, and presenting photometric features, such as morphology and colours, typical of galaxies in dense environments. The coordinates and total magnitudes of the spectroscopic NIR observational targets are detailed in Table \ref{tab1}.

\begin{figure}[H]
    \centering
    \includegraphics[width=0.8\columnwidth]{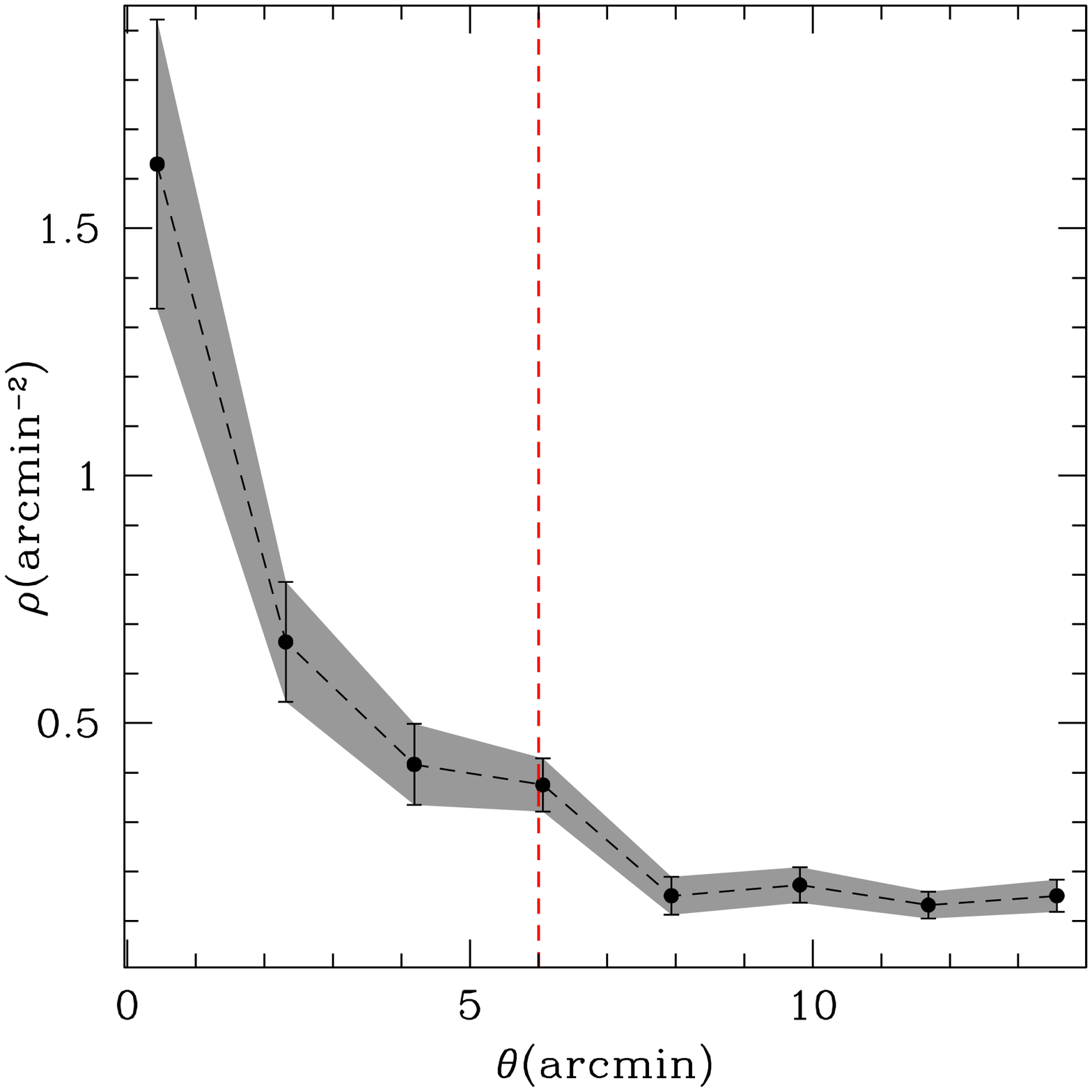}
    \includegraphics[width=1.0\columnwidth]{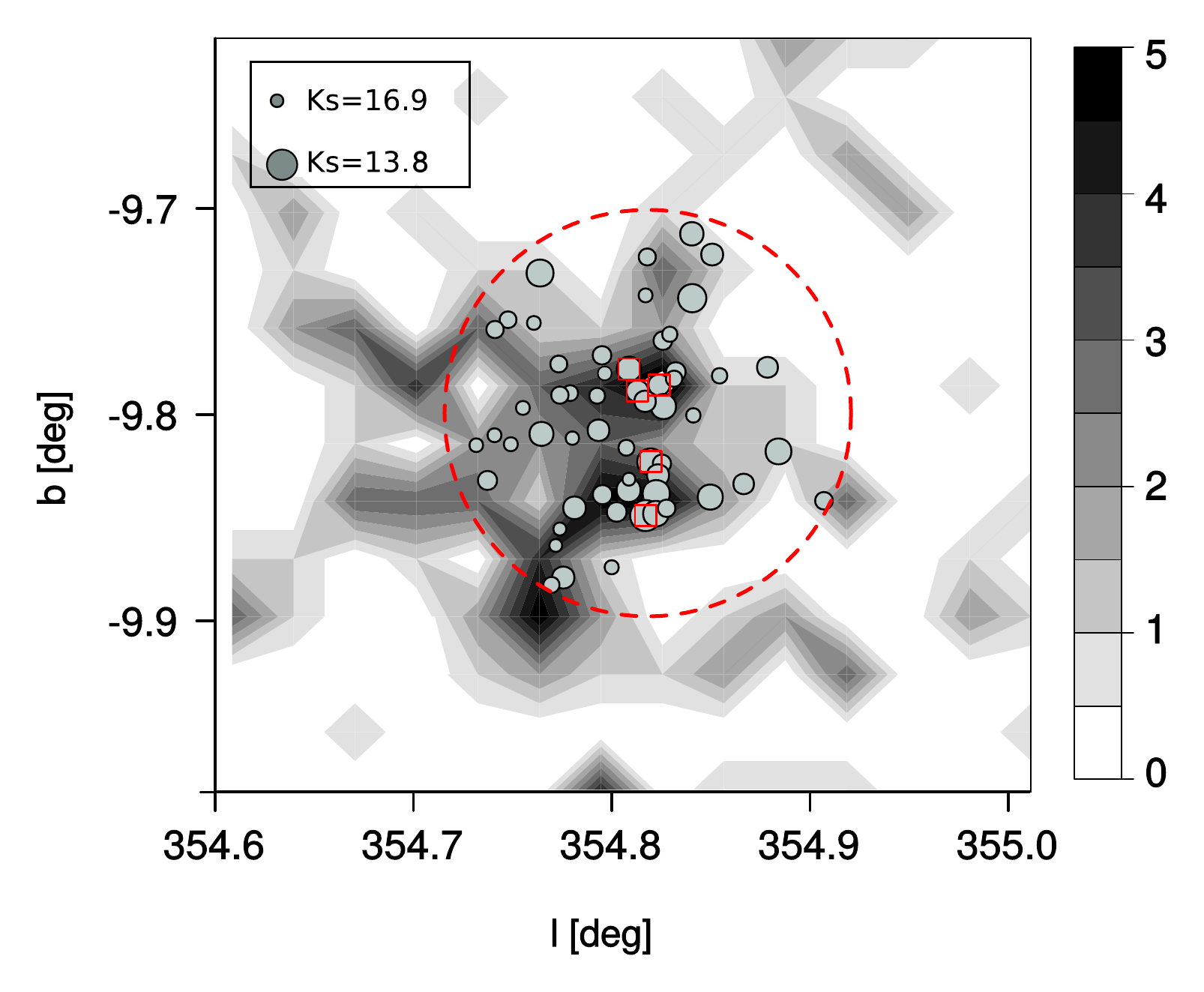}
    \caption{Top panel: Density profile as a function of the angular distance to the geometric centre of the overdensity region detected in the work of G21. The uncertainties were derived via a bootstrap resampling technique \citep{Barrow}.
    Bottom panel: Density map of visually confirmed galaxies around the central overdense region. Filled contours are colour-codded according to number counts in pixels of 2 arcmin x 2 arcmin. The dashed circle represents the six arcmin radius studied area. Grey dots indicate the positions of the 58 visually detected galaxies (dot sizes are weighted according size is proportional to $K\!s$ magnitude). The open red squares correspond to the galaxies observed with F2.}
    \label{dens}
\end{figure}

The overdensity of galaxies, behind the Milky Way, can be noticed in Fig. \ref{dens} (top panel) where the density profile estimated as a function of the angular distance from the geometric centre of this region ($l=354.82\degr$ and $b=-9.81\degr$) is presented. 
In this figure a clear excess is observed toward the region closer to the considered central position. Therefore, with the aim of confirming the existence of an overdense structure consistent with a galaxy cluster, in this paper we restricted the analysis to a smaller region around the central coordinates of the galaxy concentration considering a radius of six arcmin. This limiting value approximately corresponds to the radius where the excess becomes noticeable. In this area 58 galaxy candidates fulfil the selection criteria described in G21.
 
\begin{figure*}[htb]
    \centering
    \includegraphics[width=0.99\textwidth]{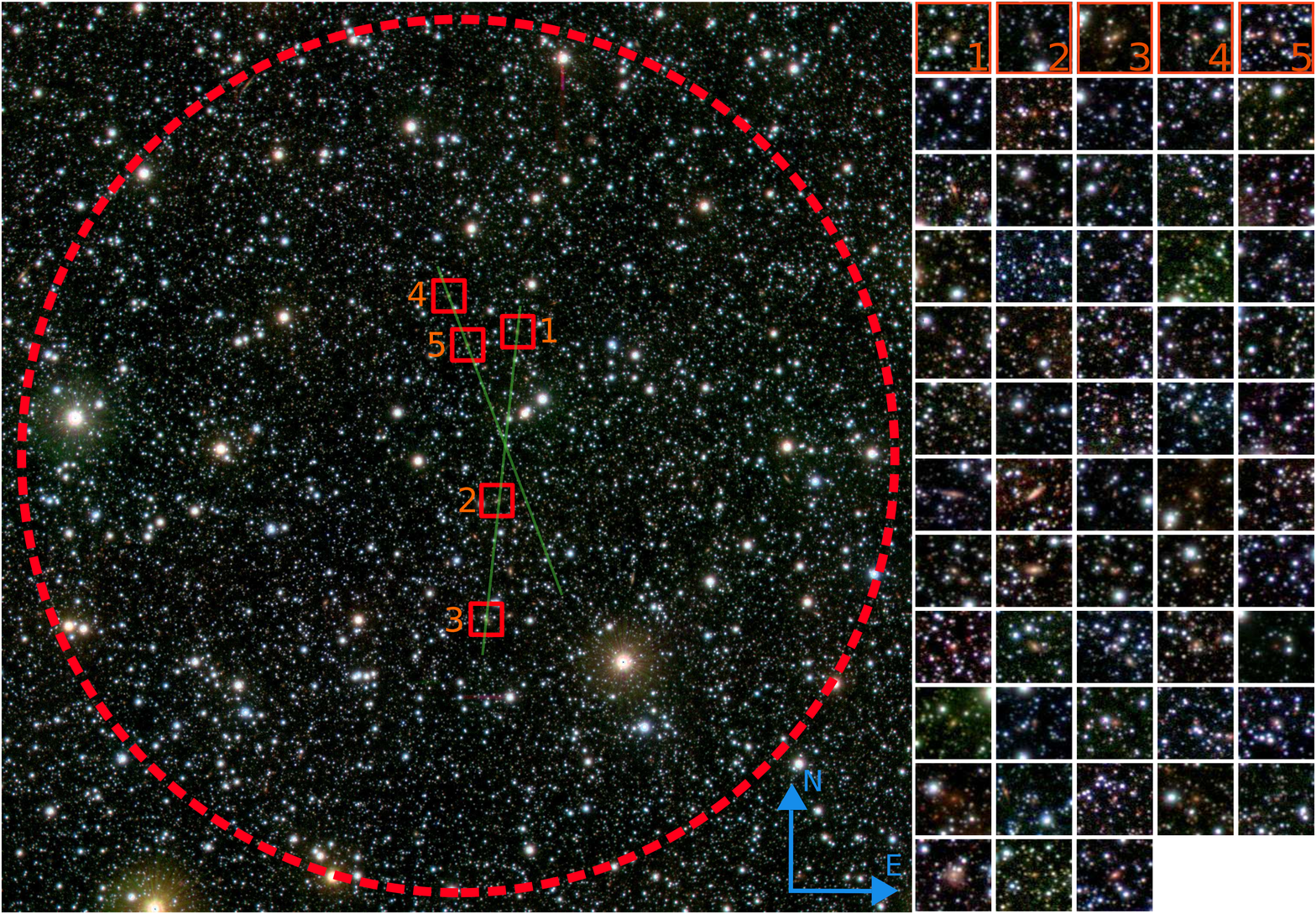}
    \caption{False-colour $Z$ (blue), $J$ (green), and $K\!s$ (red) image of a region corresponding to the galaxy group/cluster candidate. The red dashed circle delimits the six arcmin radius central area, the green lines indicate the two long-slit positions and the red squares show the five galaxies observed with F2. In the right panels we zoomed the 58 galaxy candidates within the studied area. The length of each box side is 20 arcsec.}
    \label{galaxies}
\end{figure*}

The spatial distribution of the visually confirmed galaxies in the considered area can be observed in Fig. \ref{dens} (bottom panel). From the density map is possible to detect the region with higher density of extragalactic sources. The five selected galaxies are shown in red squares. In addition in Fig. \ref{galaxies} we show a VVV false-colour multi-band ($Z$, $J$, $K\!s$) image of the six arcmin radius studied area. It is possible to appreciate the five galaxies observed with F2 (red squares). Most of the detected extragalactic sources exhibit extended morphology, typical of galaxies, and present redder colours than the foreground stellar sources as can be clearly observed for the 58 galaxy candidates zoomed in the right boxes the of Fig. \ref{galaxies}. This remarkably large number of extragalactic sources, found in a very limited region of the sky, could indicate the presence of a group or cluster of galaxies. 

For the spectroscopic observations we selected F2, which is a NIR imaging, long-slit, and recently multi-object, spectrograph at Gemini South 8.1-meter telescope, located in Cerro Pachón, Chile. This instrument offers a wavelength range of  0.9 -- 2.5 $\mu$m and a circular field of view of 6.1 arcmin on a $2048 \times 2048$ pixels HAWAII-2 detector array. F2 has a refractive all-spherical optical system providing 0.18 arcsecond pixels. 

We observed a total of five galaxies with F2 in April 2019 (Program ID: GS-2019A-Q-123). These sources were selected from the list of 58 candidates, according to their spatial distribution and brightness in order to optimise the requested time for spectroscopic observations. 
 
\begin{table*}[htb]
\centering
\caption{ID, VVV name, coordinates and total magnitudes of the spectroscopic targets properties observed with F2.}\begin{tabular}{|c c c c c c c| }
\hline
ID & VVV & RA & Dec & $J$ & $H$ & $K\!s$ \\
 &name & ($J2000.0$) & ($J2000.0$) & (Total) & (Total) & (Total)\\
\hline
\hline
  01&VVVJ181430.28-381332.7& 18:14:30.28 &-38:13:32.8 & 16.38$\pm$0.03 &  15.68$\pm$0.04 &  15.18$\pm$0.05\\
 02&VVVJ181439.84-381447.1& 18:14:39.84 &-38:14:47.1 & 16.49$\pm$0.03 &  14.83$\pm$0.03 &  14.32$\pm$0.04\\
 03&VVVJ181446.61-381537.5& 18:14:46.61 &-38:15:37.6 & 14.77$\pm$0.01 &  14.10$\pm$0.02 &  13.69$\pm$0.02\\
 04&VVVJ181426.09-381408.0& 18:14:26.09 &-38:14:08.0 & 16.13$\pm$0.03 &  15.38$\pm$0.04 &  14.82$\pm$0.04\\
 05&VVVJ181429.59-381412.3& 18:14:29.59 &-38:14:12.4 & 16.40$\pm$0.03 &  15.71$\pm$0.04 &  15.09$\pm$0.05\\
\hline  
\end{tabular}
\label{tab1}
\end{table*}

The observations were made by using HK (1.2-2.4$~\mu$m) band grism. Taking into account the VISTA telescope scale and the half-light radius calculated with SExtractor for the five galaxy candidates, we estimate the apparent width of these objects in approximately 1.1 arcsec.
In this way we selected the 6 pixels longslit width in order to collect the galaxy signal but avoiding the light of the stars on the field. In this configuration we obtained a spectral resolution close to 1400. 
The observations were performed with an airmass lower than 1.2 and no cloud coverage. The average seeing during the observing run was estimated at 0.9 arcsec.
For each target we observed 12 x 110 sec (0.37 hr) of exposure time and the observational sky conditions were optimal allowing to reach a S/N$\sim$70 for all galaxies in the spectral region between 2,1 $\mu$m and 2,2 $\mu$m. With the first slit position we observed three galaxies with $K\!s$ magnitudes of 15.18, 14.32 and 13.69, respectively. The angular projected distance between the first and second galaxy is 2.25 arcmin and between the first and third is 3.81. We located the longslit at a position angle of 122° from E to N. At the second slit position we observed two group/cluster galaxy candidates with magnitudes $K\!s = 15.09$ and $K\!s =14.82$, respectively. The distance between galaxies is 0.42 arcmin. In this case the position angle of the longslit was 97° from E to N (see for instance green lines in Fig. \ref{galaxies}). In order to provide telluric standards at similar air masses we observed HIP~92688, an AO V star. 

The spectra were reduced with Gemini IRAF package Version 1.14. As part of the basic data reduction steps we created the necessary dark images and the normalised flat field. Then, we reduced the arc and determined the wavelength solution. We reduced and combined the telluric data and applied the wavelength calibration to extract the telluric spectrum. Next, we reduced and combined the science data. These steps were carried out with \textit{nsreduce}, \textit{nscombine}, \textit{nsfitcoords}, \textit{nstransform} and \textit{nsextract} routines. Afterwards, the wavelength calibration was applied to the science data in order to extract the science spectrum, using suitable apertures to reduce the noise in each one in order to reach the best S/N. Finally, we applied the telluric correction.

\section{Results}
\label{resN}

The knowledge obtained from spectroscopic observation using F2, from Gemini, can be reinforced analysing the photometric information available from VVV survey. In this section we describe both, spectroscopic and photometric results.

\subsection{Spectroscopic redshift estimation}
\label{SR}

\begin{figure*}[htb]
    \centering
    \includegraphics[width=0.7\textwidth]{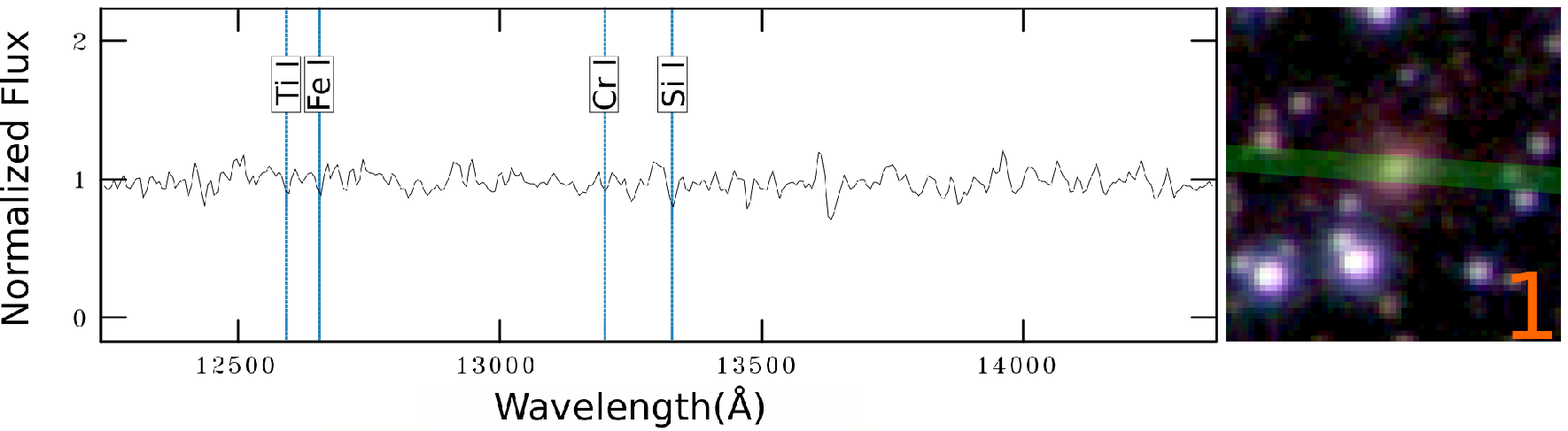}
    \includegraphics[width=0.7\textwidth]{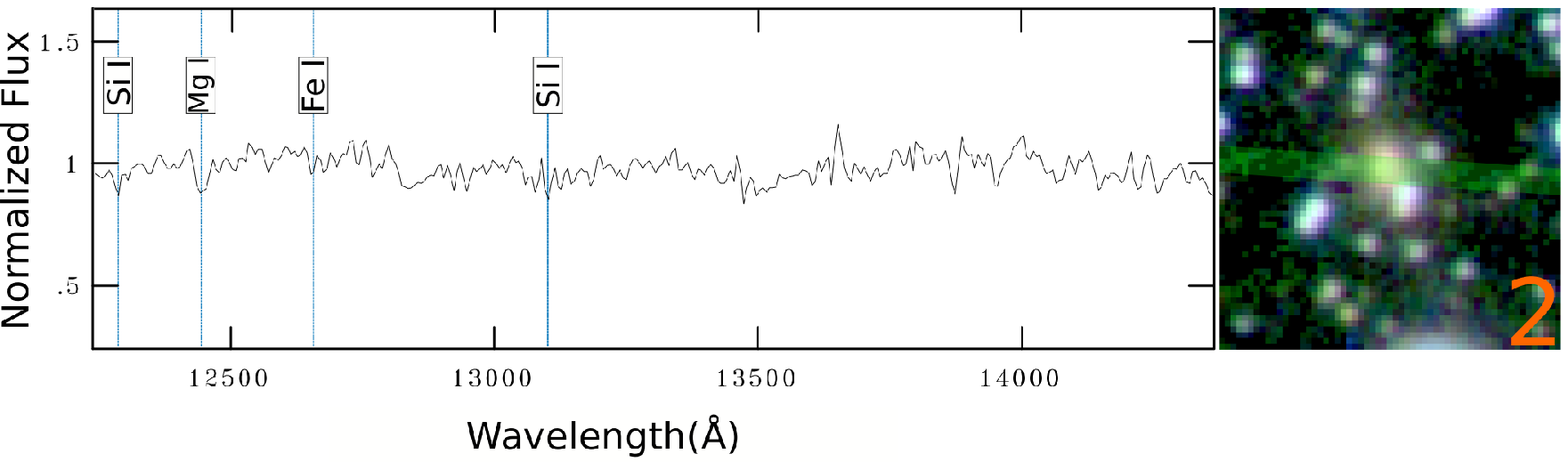}
    \includegraphics[width=0.7\textwidth]{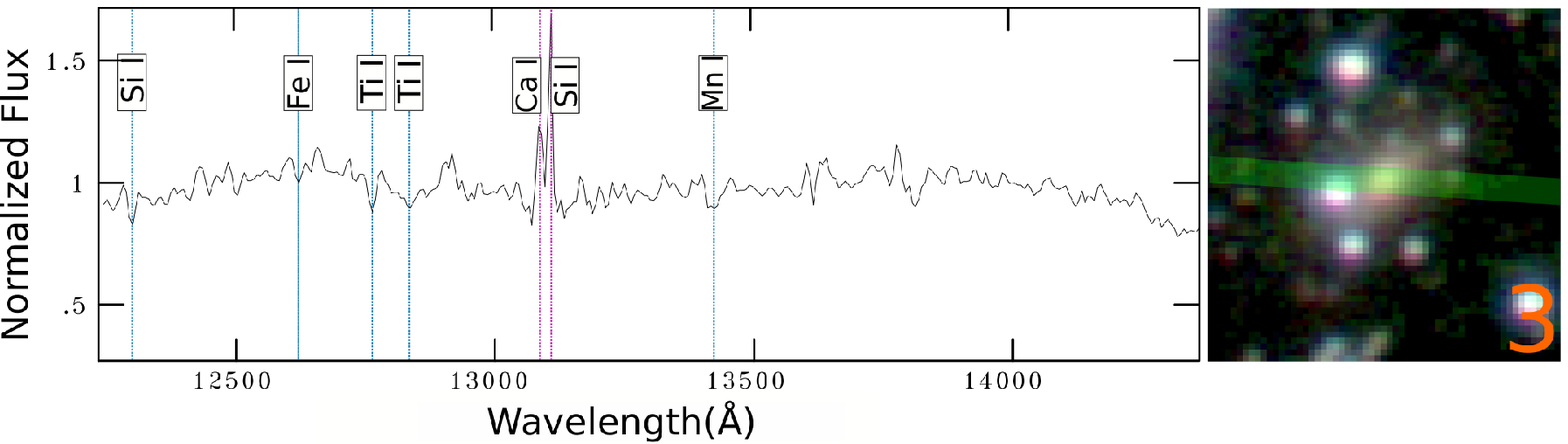}
    \includegraphics[width=0.7\textwidth]{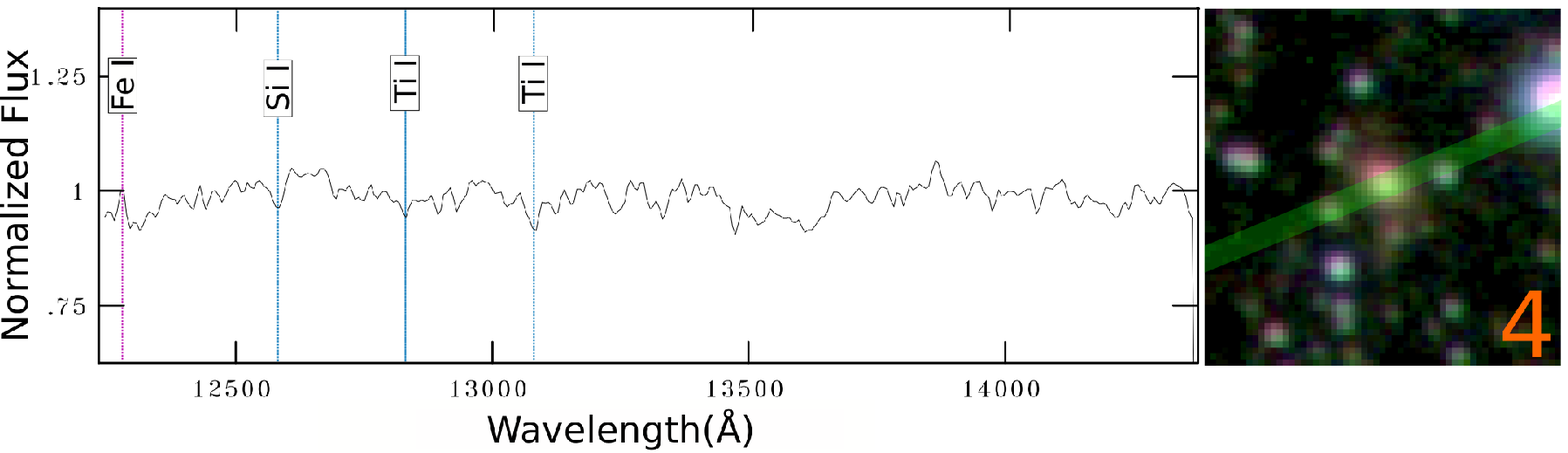}
    \includegraphics[width=0.7\textwidth]{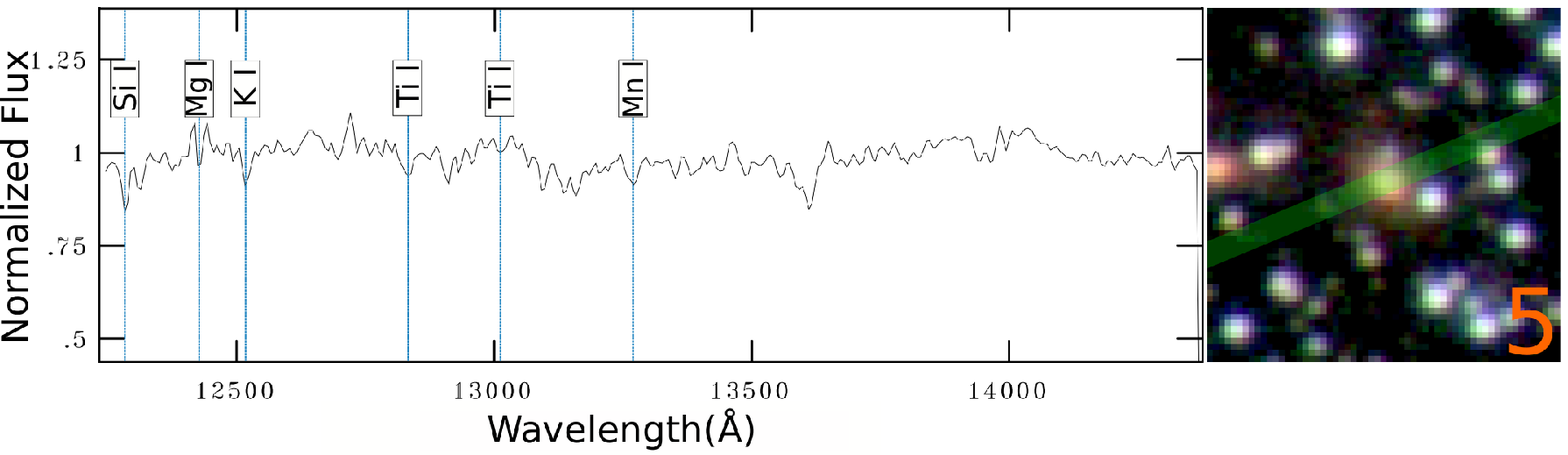}
    \caption{Left: Final reduced and redshift-corrected NIR spectra. Right: RGB false colour image  of the observed galaxies as in Fig. \ref{galaxies}. The length of each box side is 20 arcsec. The green lines indicate the slit position.}
    \label{spec1}
\end{figure*}

The galaxies inhabiting high density regions have identifiable characteristics such as, e.g., absorption lines and red colours. In Fig. \ref{spec1} we present the NIR spectral features and the detected lines.
The RGB images from these galaxies show their bulge-type morphology and red colours, as can be expected for galaxies within galaxy clusters following the morphology-density relation from \cite{dre80}.

The observed spectra clearly present absorption features. Using as a first approach the photometric redshift estimates (see \S~\ref{PR}), these characteristics are located in the NIR spectral region, which is rich in absorption components \citep[see][]{riffel11, riffel15, riffel19}. To correctly identify these features we have cross correlated the observed spectra with a set of stellar spectra, using the {\sc starlight} code \citep{cidfernandes04, cidfernandes05, asari07, cidfernandes18}. The procedures we have followed are described in \citet{riffel09, riffel15, riffel22}, including, for example, the methods used to handle extinction, emission lines and differences in spectral resolution. Briefly, {\sc starlight} fits an observed spectrum $O_{\lambda}$ with a combination, in different proportions, of $N_{*}$ spectral base elements, solving the equation:

\begin{equation}
M_{\lambda}=M_{\lambda 0}\left[\sum_{j=1}^{N_{*}}x_j\,b_{j,\lambda}\,r_{\lambda} \right] \otimes G(v_{*},\sigma_{*}),
\end{equation}

where $M_{\lambda}$ is a model spectrum, $b_{j,\lambda}\,r_{\lambda}$ is the reddened spectrum of the $j$th $N_{*}$ normalised at $\lambda_0$; $r_{\lambda}=10^{-0.4(A_{\lambda}-A_{\lambda 0})}$ is the reddening term; $M_{\lambda 0}$ is the theoretical flux at the normalisation wavelength; $\vec{x}$ is the population vector, and $G(v_{*},\sigma_{*})$ is the Gaussian distribution used to model the line-of-sight stellar motions, which is centred at velocity $v_{*}$ with dispersion  $\sigma_{*}$. The final fit is carried out searching for the minimum of the equation:

\begin{equation}
\chi^2 = \sum_{\lambda}[(O_{\lambda}-M_{\lambda})w_{\lambda}]^2,
\label{equchi}
\end{equation}

where emission lines and spurious features are masked out by fixing $w_{\lambda}$=0 (normally, $w_{\lambda} = 1/e_{\lambda}$, with $e_{\lambda}$ being the uncertainty in $F_{\lambda}$). The quality of the fit is accessed by $\chi^2_{Red}$ which is the $\chi^2$ given by equation~\ref{equchi} divided by the number of points used in the fit and by $adev = |O_{\lambda} - M_{\lambda}|/O_{\lambda}$, which is the percentage mean deviation over all fitted pixels. For a detailed description of {\sc starlight} see its manual\footnote{http://astro.ufsc.br/starlight/}. 

Since we are interested in identifying the stellar features to define the set of spectra used by the code to fit the underlying continuum we have followed the approach of \citet{riffel15} and used their {\it stars approach}, which is a base composed of all the 210 dereddened stars (spectral types F--S/C, F being the hottest available) in the IRTF spectral library \citep{rayner09,cushing05}.
After this procedure, we have been able to recognised absorption and emission lines considering \cite{rayner09}. The spectral lines detected in the majority of galaxies are FeI, SiI, TiI, and MgI. Besides that, CrI, KI, CaI, and MnI lines have been detected in a some cases. 

The NIR detected lines for every individual galaxy, shown in Fig \ref{spec1} (left panels), are summarised as follow:
\begin{itemize}
\item ID 01 - VVVJ181430.28-381332.7: presents TiI (1.259$\mu$m), FeI (1.265$\mu$m), CrI (1.320$\mu$m), and SiI (1.332$\mu$m) lines in its spectrum, all them are absorption lines. From these lines the estimated redshift is $z=0.225 \pm 0.003$.
\item ID 02 - VVVJ181439.84-381447.1: shows SiI (1.228$\mu$m), MgI (1.244$\mu$m), FeI (1.257$\mu$m), and SiI (1.310$\mu$m) in absorption, with a measured redshift $z=0.226 \pm 0.003$. 
\item ID 03 - VVVJ181446.61-381537.5: presents CaI (1.308$\mu$m) and SiI (1.310$\mu$m) in emission, and SiI (1.229$\mu$m), FeI (1.262$\mu$m), TiI (1.283$\mu$m), and MnI (1.342$\mu$m) in absorption. The measured redshift is $z=0.225 \pm 0.004$.
\item ID 04 - VVVJ181426.09-381408.0 shows FeI (1.227$\mu$m) in emission, and SiI (1.258$\mu$m) and TiI (1.283$\mu$m, 1.308$\mu$m) in absorption,  with a redshift $z=0.225 \pm 0.002$.
\item ID 05 - VVVJ181429.59-381412.3 presents SiI (1.228$\mu$m), MgI (1.242$\mu$m), KI (1.252$\mu$m), TiI (1.283$\mu$m, 1.302$\mu$m) and MnI (1.327$\mu$m) in absorption, with a redshift $z=0.225 \pm 0.004$.
\end{itemize}
The measured redshifts of the five observed galaxies are pretty similar suggesting that they belong to a common system of galaxies with an estimated mean redshift of $z=0.225 \pm 0.014$. This finding allows to speculate about the existence of a galaxy cluster behind the galactic bulge. In the following section the photometric information of these galaxies, and the remaining ones in the studied region, could give light to this finding.

\subsection{Red sequence}
\label{RS}
Galaxy colours provide strong information about the evolutionary processes they have been subjected to, and therefore indicative of the environment where they reside. The colour–magnitude diagrams of the galaxy clusters, known as the red-sequence \citep{gla00} contain a well-defined, highly regular population of early-type galaxies, and have also been used by several authors to successfully identify clusters of galaxies (e.g. \cite{gla00}; \cite{LopezCruz04}; \cite{Soech06}).  

\begin{figure*}[htb]
    \centering
    \includegraphics[width=0.99\textwidth]{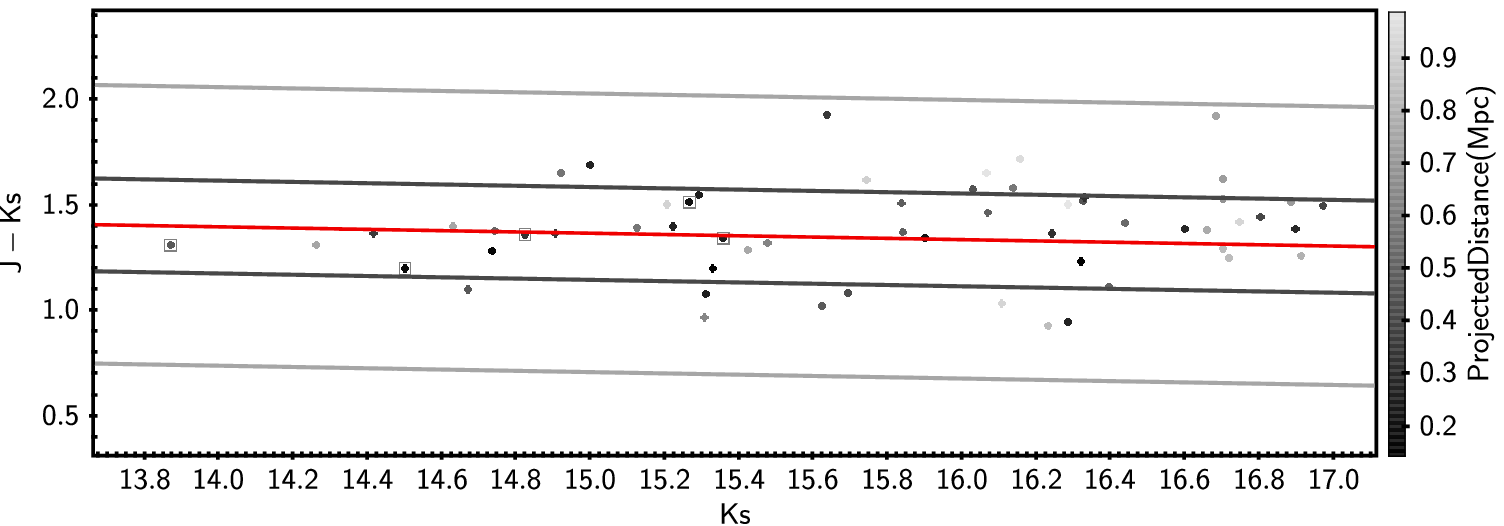}
    \caption{Colour-magnitude diagram $J-K\!s$ versus $K\!s$. The points are colour--coded according to the projected distance at the centre of the overdensity zone. The red sequence model is shown as a red line, the dark grey lines represents $\pm 1 \sigma$ around the model and the light grey lines represents $\pm 3 \sigma$. The black squares represents the five galaxies observed with F2.}
    \label{RS_plot}
\end{figure*}

In order to determine the nature of this overdense region we use the VVV photometric information (as described in G21) of the galaxies within six arcmin radius from the estimated position of the overdensity to analyse the colour-magnitude relation. To further strengthen the identification of this region as a galaxy cluster we used the model of \cite{stott2009} calculated from semi-analytical model of \cite{bower2006}, considering the measured spectroscopic redshift $z=0.225 \pm 0.014$, obtaining a red-sequence model as shown in the colour–magnitude diagram of Fig. \ref{RS_plot}. Then, we performed $\chi-squared$ goodness-of-the-fit test finding a $p>0.05$ value. Therefore, we can not reject the null hypothesis that this red-sequence model suitably fits our data.

Fig. \ref{RS_plot} shows that all the galaxies in the 6 arcmin radius area are found within $\pm 3 \sigma$ around the red-sequence model. Further, by restricting the selection to $\pm 1 \sigma$ around the linear model we found 40 galaxies ($\sim69$ \%). This finding suggests that galaxies from this structure can reproduce a well-defined red-sequence corresponding to a galaxy cluster at the spectroscopic redshift estimated in Sec. \ref{SR}. This result is consistent with that found by \cite{bar19} and \cite{gal22} for other galaxy clusters studied using the VVV survey.

\subsection{Photometric redshift} 
\label{PR}
In order to investigate the redshift distribution of the 58 galaxy candidates in the area under study we estimate photometric redshift by running the software EAZY \citep{bra08}. We consider the default set of parameters using all templates simultaneously, v1.0 template error function and the K-extended prior. To calculate the limiting redshift we consider that VVV photometry is three magnitude deeper than 2MASS \citep{min10}. Therefore taking into account the limiting redshift of 2MASS extended sources around $z=0.2$ and $K\!s$ magnitude limit $K\!s=15$ \citep{Bilicki2014} we can calculate an absolute magnitude limit $\rm M_{K\!s}\sim-25$ for 2MASS extended sources. If we take the limiting magnitude of our candidates $K\!s=17$, we can observe a galaxy with absolute magnitude $\rm M_{K\!s}\sim-25.5$ up to $z \sim0.45$, therefore we allowed photometric redshift solutions in the range $0<z<0.45$ with a step of 0.01.

\begin{figure}[htb]
    \centering
    \includegraphics[width=1.0\columnwidth]{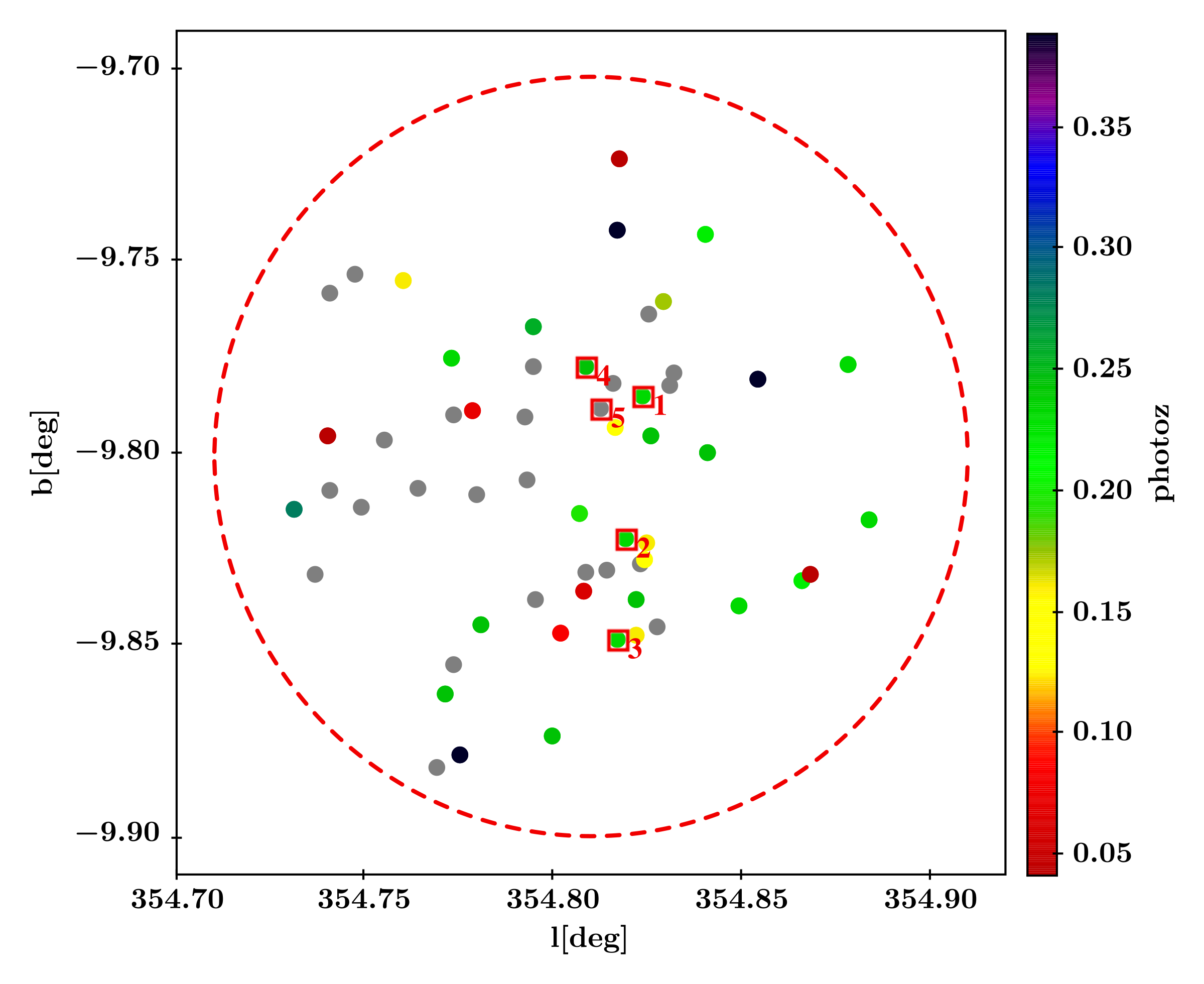}
    \caption{ Sky distribution of the 58 extended sources in the overdensity zone, colour--coded according to the obtained photometric redshift $photoz$, grey dots represent objects with unreliable estimates and the open red squares are the galaxies observed with F2. The dashed circle represents the six arcmin area under study.}
    \label{radeczpaux}
\end{figure}

\begin{figure}[htb]
    \centering
       \includegraphics[width=0.9\columnwidth]{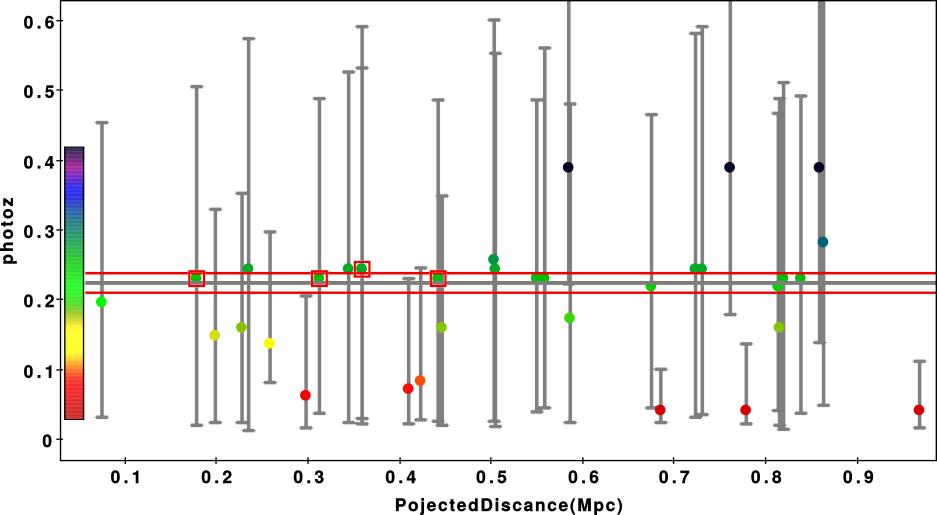}
        \includegraphics[width=0.9\columnwidth]{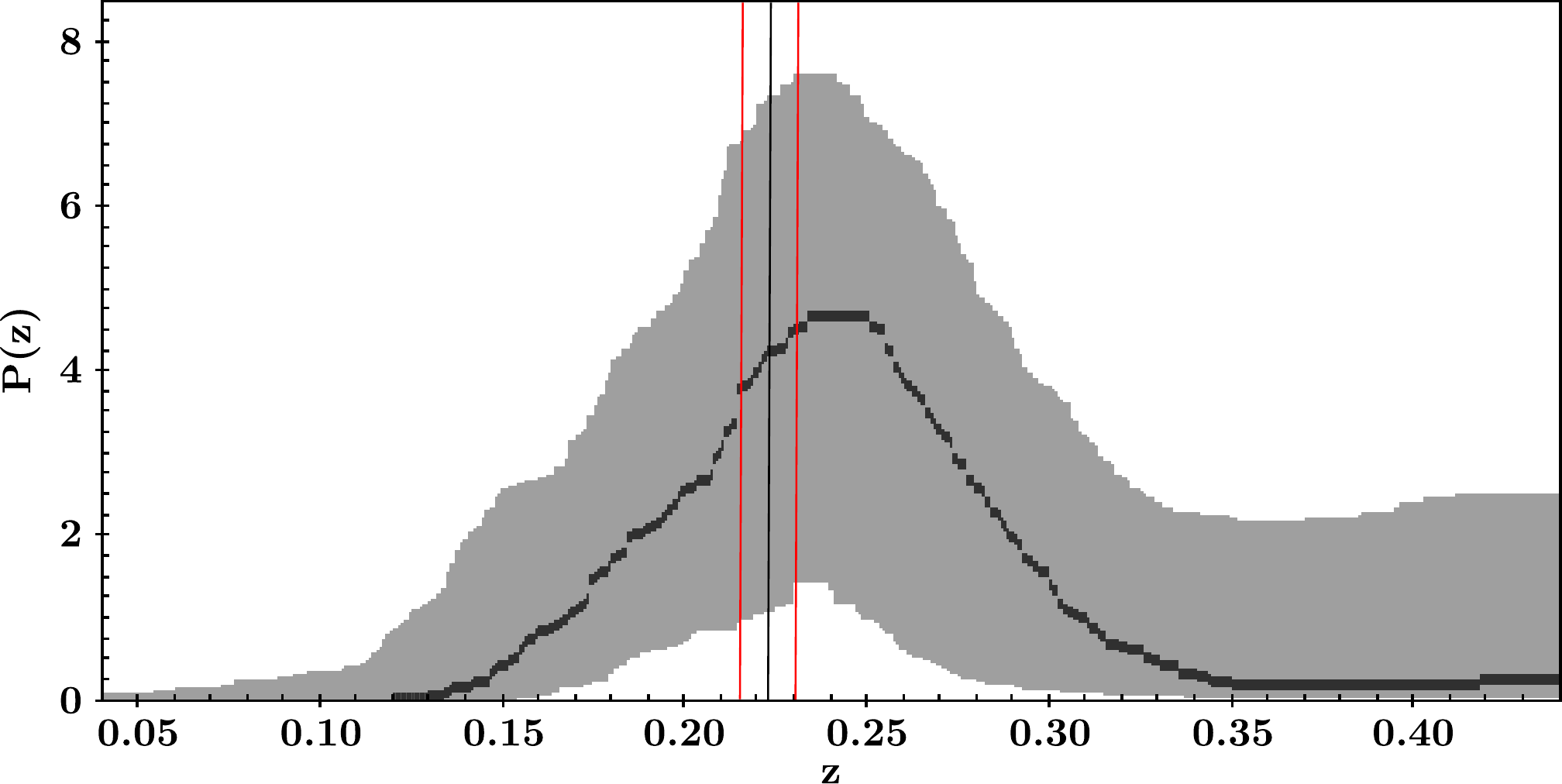}
    \caption{Top: Distribution of $photoz$ as a function of the projected distance to the centre. The error bars correspond to $1\sigma$ individual uncertainties in the photometric redshift estimates. F2 observed galaxies are marked as red squares.  Bottom: Average probability distribution function of the 34 galaxies with reliable photometric redshifts in our sample. The line corresponds to the mean values and the shaded region to 1$\sigma$ from the mean.The lines show the obtained spectroscopic redshift (grey) and its associated errors (red).}
    \label{Fpz}
\end{figure}

We built a photometric input catalogue that contains photometric fluxes and uncertainties observed in YJHKs VVV filters, requiring a minimum of three fluxes to perform the fit. Also we require that \texttt{peak-prob}$>$0.9  to prevent unreliable photometric redshift fits. Under these constrains we can estimate photometric redshifts for 34 galaxy candidates,  obtaining a mean redshift $photoz=0.21 \pm 0.08$. 

In Figure \ref{radeczpaux} we show the spatial distribution of galaxies in our sample, colour--coded according to the obtained photometric redshift. From this figure we can appreciate the good agreement of the galaxy photometric redshifts in the area under study. Also, there are some galaxies with estimated photometric redshift differing significantly  from the obtained spectroscopic redshift (six with $photoz\sim$0.05 and three with $photoz\sim$0.4). To study contaminant galaxies we plot in the top panel of Fig. \ref{Fpz} the distribution of $photoz$ as a function of the projected distance to the centre. We include, in this plot, individual uncertainties in the photometric redshift estimation and show the obtained spectroscopic redshift and its associated errors.  As can be appreciated three of the most distant galaxies with $photoz\sim$0.05 can be considered as interlopers because they have a difference with the spectroscopic redshift larger than 3$\sigma$. The errors associated to galaxies with $photoz\sim$0.4 are too large to draw any conclusions.

To strengthen our result we carry out an inspection of the probability distribution functions of galaxies with reliable photometric redshift. In the bottom panel of Fig. \ref{Fpz} we show the average probability distribution function for these 34 objects finding that the resulting distribution is well behaved. We also plot the obtained spectroscopic redshift and its associated error (vertical red and grey lines, respectively). From the figure it can also be appreciated that the peak of the  average probability distribution function is slightly moved toward higher redshifts with respect to the measured spectroscopic redshifts, being the difference lower than within 3$\sigma$.

\section{Discussion and summary}
\label{sum}

Considering the high contamination by dust, gas and stellar objects in the Milky Way central region, it is well known the lack of information about the extragalactic objects located behind this zone. To overcome this constraint, we analysed NIR images with the aim to reduce the contamination and highlighting the light coming from external galaxies. In this work we present an analysis of an extragalactic overdense region located in the tile $b204$ of the VVV survey.

In this study we restricted the analysis to an area of six arcmin around a galaxy concentration located at $l=354.82\degr$ and $b=-9.81\degr$. 
In this area 58 galaxy candidates can be observed fulfilling the selection criteria described in G21. From this sample we carefully selected five galaxies to obtain NIR spectra taking into account the magnitude and spatial distribution in order to optimise the requested time for spectroscopic observations. In this way five spectra were taken with Flamingos 2 long-slit spectrograph at Gemini South 8.1-meter telescope. 

We found that all the spectra display absorption features, in agreement with expected characteristics considering the morphology of the observed galaxies. 
Also we carried out a stellar population synthesis on the mean spectrum using the {\sc starlight} spectral synthesis code. In this way we found out that most galaxies show the continua dominated by stellar absorption features. The most abundant elements are Fe I, Si I, and Ti I. The spectroscopic redshift was calculated for every single galaxy, considering all lines identified in each spectrum, finding a mean redshift $z=0.225\pm 0.014 $. Taking into account this result, the six  arcmin radius used for the analysis performed in this paper corresponds to $\approx$ $1 \mpc$, which represents a typical galaxy cluster radius, located at this redshift.

In order to reinforce our result, complementary techniques such as red sequence and photometric redshift were applied. In this way, we analysed the NIR colour-magnitude diagram, considering the $K\!s$ magnitude and the J-Ks colours. 
Then, the red sequence model was performed following \cite{stott2009} for a galaxy cluster at measured spectroscopic redshift $z=0.225\pm 0.014$ finding 58 and 40 galaxies ($\sim$100 \% and $\sim$69 \%), within $\pm 3 \sigma$ and $\pm 1 \sigma$ around the red-sequence linear model, respectively.

Furthermore, we estimate the photometric redshift of the galaxy candidates by running the software EAZY. We applied constraints requiring a minimum of three fluxes and \texttt{peak-prob}$>$0.9 to ensure reliable fits. In this sense the final sample has 34 galaxy candidates with reliable photometric redshift estimations with a mean value $photoz=0.21 \pm 0.08$. Also, we made an inspection of the probability distribution functions of galaxies with reliable photometric redshifts, finding that the resulting average distribution is well behaved with a mean and dispersion in good agreement with the average $photoz$.

Finally, considering the spectroscopic redshifts measurement of five galaxy members we performed the estimation of the line-of-sight velocity dispersion of the cluster candidate using the gapper estimator described by \cite{Beers90}. Thus, we obtained for the velocity dispersion a value of $\sigma \approx 400 \kms$. In addition,  we calculated the virial radius and the virial mass of the galaxy cluster candidate following \cite{MZ05} finding $R_{vir} \approx  1.19 \mpc$ and $M_{vir} \approx 4.43 \times 10^{13} M_{\sun}$.
Although a high amount of spectroscopic observations will allow to obtain accurate values for the cluster parameters, these results are consistent with that expected for rich galaxy groups \citep{dom02, MZ05} or a galaxy cluster with a virial mass according to the redshift range $z=0.1-0.4$ \citep{wies}.

The agreement of the redshifts obtained from the three different methods 
and the estimated cluster parameters allow to confirm the nature of this structure as a galaxy cluster at $z=0.225 \pm 0.014$, named VVVGCl-B J181435-381432, unveiling a new extragalactic system that was hidden behind the Milky Way bulge.
\begin{acknowledgements}

This work was partially supported by the Consejo Nacional de Investigaciones Cient\'{\i}ficas y T\'ecnicas and the Secretar\'{\i}a de Ciencia y T\'ecnica de la Universidad Nacional de San Juan. The authors gratefully acknowledge data from the ESO Public Survey program ID 179.B-2002 taken with the VISTA telescope, and products from the Cambridge Astronomical Survey Unit (CASU). 
This paper is based on observations obtained at the international Gemini Observatory, a program of NSF’s NOIRLab, which is managed by the Association of Universities for Research in Astronomy (AURA) under a cooperative agreement with the National Science Foundation on behalf of the Gemini Observatory partnership: the National Science Foundation (United States), National Research Council (Canada), Agencia Nacional de Investigaci\'{o}n y Desarrollo (Chile), Ministerio de Ciencia, Tecnolog\'{i}a e Innovaci\'{o}n (Argentina), Minist\'{e}rio da Ci\^{e}ncia, Tecnologia, Inova\c{c}\~{o}es e Comunica\c{c}\~{o}es (Brazil), and Korea Astronomy and Space Science Institute (Republic of Korea).
RR thanks to Conselho Nacional de Desenvolvimento Cient\'{i}fico e Tecnol\'ogico ( CNPq, Proj. 311223/2020-6, 304927/2017-1 and 400352/2016-8), Funda\c{c}\~ao de amparo \`{a} pesquisa do Rio Grande do Sul (FAPERGS, Proj. 16/2551-0000251-7 and 19/1750-2), Coordena\c{c}\~ao de Aperfei\c{c}oamento de Pessoal de N\'{i}vel Superior (CAPES, Proj. 0001).
D.M. gratefully acknowledges support by the ANID BASAL projects ACE210002 and FB210003 and by Fondecyt Project No. 1220724.
This research made use of \texttt{Astropy},\footnote{http://www.astropy.org} a community-developed core \texttt{Python} package for Astronomy \citep{astropy:2013, astropy:2018}. 
\end{acknowledgements}

\bibliography{references}

\end{document}